# Journey of Cryptocurrency in India
## In View of Financial Budget 2022-23

**Varun Shukla, Manoj Kumar Misra, Atul Chaturvedi**
Pranveer Singh Institute of Technology (PSIT), Kanpur, India

**Abstract**: Recently, Indian Finance minister Nirmala Sitharaman announced in Union budget 2022-23 that Indian government will put 30% tax (the highest tax slab in India) on income generated from cryptocurrencies. Big financial institutions, experts and academicians have different opinions in this regard. They claim that it would be the end of cryptocurrency market in India or it would be possible that RBI (Reserve Bank of India) may launch its own crypto or digital currency. So in this context, in this article, the journey and future aspects of cryptocurrency in India are discussed and we hope that it will be a reference for further research and discussion in this area.

1. **Cryptocurrency: Definition and types [1-3]**
   - **Definition**: Cryptocurrency can be seen as a digital/virtual currency and cryptography provides very strong security to it and it is next to impossible to do double spending or counterfeiting with cryptocurrency. A cryptocurrency follows decentralized network which is based on Blockchain concept (Distributed Ledger). The exchange of cryptocurrency occurs between consent parties without any involvement of brokers or regulating agencies.
   - **Types of cryptocurrencies**: Only major cryptocurrencies are given below in **table 1** in decreasing order of their approximate market value (as it is subject to change very frequently) for the ready reference of readers of this article.

| S.N. | Name | Market value (in billion dollars) | Current price (in dollars) |
|---|---|---|---|
| 1 | Bitcoin | 825 | 36,729.32 |
| 2 | Ethereum | 388 | 2,624.64 |
| 3 | Binance coin | 80 | 365.31 |
| 4 | Cardano | 65 | 1.04 |
| 5 | Doge coin | 62 | 0.14 |
| 6 | Thther | 57 | 1.00 |
| 7 | XRP | 50 | 0.60 |
| 8 | Polkadot | 31 | 18.43 |
| 9 | Internet Computer | 24 | 20.43 |
| 10 | Bitcoin Cash | 21 | 275.45 |

**Table 1**: Showing market value and current price of major cryptocurrencies

## 2. Cryptocurrency in India So Far [4-7]

- **During 2013-17**: The duration 2013-17 can be seen as a starting of cryptocurrency trend in India. Doubts, ambiguities and confusions were very high for cryptocurrencies development in India during this time. In 2013, RBI has warned public about cryptocurrencies. Here the term "public" means consumers, holders, trading persons/agencies etc. RBI has also stated that it is observing all the cryptocurrency related developments very closely including Bitcoins (very popular one) and other cryptocurrencies (Altcoins-An Altcoin is an alternative digital currency to Bitcoin). In February 2017, RBI has warned the public again and in last quarter of 2017, RBI has issued a clear cut warning that "virtual currencies/cryptocurrencies are not a legal tender in India". Due to some PILs (Public Interest Litigations) filed in court as a reaction of RBI's warning, the government of India has formed a committee to look various issues related to cryptocurrency and tell about the required actions in future. So as a result, no ban on virtual currencies took place in this period.
- **2018**: It is very important to mention that in April 2018, the committee appointed by finance ministry of India, has drafted a bill regarding cryptocurrencies but "was not in the favour of ban".
- **2019 (Banning of cryptocurrency bill):** Some major points are given below

- Trading, mining, holding or transferring/use of cryptocurrencies is subject to punishment in India with a financial penalty or/and imprisonment up to 10 years.
- Any holder/user/person must declare/dispose of all cryptocurrencies in his/her possession within the time span of 90 days with effect from the publish date of this act.
- The processes/technology of cryptocurrencies can be utilized for research/development or teaching and academic purposes.
- RBI may launch digital rupee as a legal tender in India in future.
- The government of India may give relaxation in certain trading activities in public interest (if necessary).

- **2020 (Setback for RBI)**: In March 2020, the supreme court of India has given a set back to RBI by removing the ban on cryptocurrencies imposed by RBI.

  - **Judgement**: A bench including justices Rohinton Nariman, Aniruddha Bose and V.Ramasubramanian have passed a 180 page verdict claiming that: RBI has not mentioned any point regarding regulations of virtual currencies till date by nationalised banks/commercial banks/other financial institutions etc and also RBI did not yet mentioned any adverse effect directly or indirectly, completely/partially due to the exchange of virtual currencies. Justice Ramasubramanian who headed the bench said that the RBI stand was very "disproportionate". The Supreme Court has also indicated the failure of government of India that even after several bills and committees, government is failed to introduce any legal digital Indian rupee.

- **2021**: A high level Inter-Ministerial Committee (IMC) was formed (having secretary of economic affairs as chairman of the committee) and the job of the committee is to study various issues of cryptocurrencies and suggestions for future actions. All this has been declared by finance minister Nirmala Sitharaman on February 2021 in Rajya Sabha. In the same context and continuation, minister of state for finance Anurag Thakur has also announced in the parliament that the government is very determined to present a bill on cryptocurrencies. Finance minister Nirmala Sitharaman has also indicated that the government wants to enhance research and innovation in crypto-related areas with an open mindset. In November 2021, the standing committee on finance, headed by Jayant Sinha hold a meeting with various representatives of crypto-exchanges in

India along with Blockchain and Crypto Assets Council (BACC) and concluded that it will be unfair to ban cryptocurrencies in India but it should be regulated. Just after that, RBI governor Shaktikanta Das mentioned that cryptocurrencies can be a threat to the financial system because they are unregulated and discussed the keen intention of RBI to launch its own digital currency (as a legal tender of course). The finance minister Nirmala Sitharaman stated in Rajya Sabha that the government has not taken any step towards banning of cryptocurrencies advertisements in India but government will spread awareness on cryptocurrencies through RBI and SEBI (Securities and Exchange Board of India).

3. **Impact of Union Budget 2022-23 on Cryptocurrencies in India [8]**
    - ➢ The government of India has clearly mentioned in union budget 2022-23 that-the transfer of any virtual currency/cryptocurrency asset will be subject to 30% tax deduction.
    - ➢ No loss in the transaction will be permitted to be carried forward.
    - ➢ Gifts in the form of virtual assets/cryptocurrencies will be taxed in the hands of the receiver.
    - ➢ A Central Bank Digital Currency (CBDC) by utilizing the concept of Blockchain will be issued by RBI by the year 2023.
    - ➢ A tax of 1% will be deducted at source for the payments made on the transfer of digital assets.
   - **Impact**: This clear cut announcement of Indian government may recognize cryptocurrencies as a legitimate asset and the corresponding trading as a legal activity. The clarity on tax slab clarifies doubts and may increase the industry size.
   - **Bad for investors**: No carry forward losses will be a setback for investors as cryptocurrencies are highly volatile. This fear will always discourage investors (specifically retail investors) from trading in cryptocurrencies. The high tax slab will cut down the net profit of investors and with effect from 1st April 2023, the 115BBH provisions on income generated from virtual currencies will be in force.
   - **Dilemma**: Imposing tax on cryptocurrencies does not completely and explicitly declare cryptocurrencies legal because income tax in India is subject to assets not on the method/manner of acquiring those assets.

- **Future**: From previous history and recent tax declarations, it can be said that it is very unlikely that government of India will introduce a fresh bill to declare cryptocurrencies illegal.

**Conclusion**: It can be concluded from the above discussion that the journey of cryptocurrency is not too long in India but it has seen many ups and downs in this short span. The banning of cryptocurrencies bill in 2019 and Supreme Court verdict in 2020 are the key issues. Cryptocurrencies have a high potential and recently after union budget of 2022-23 (presented in 1st February 2022), Indians have once again started talking about it. It will be very interesting to see that after 30% tax impositions, how investors react about cryptocurrencies in India. The launch and features of RBI's-future digital currency will also be very important. After the union budget 2022-23, investors are started saying that India is following China by giving sole authority to RBI to launch and promote digital currencies. If government of India will present fresh bill on cryptocurrency, it will be very interesting to see the nature and regulations of it. Apart from all the facts and predictions, one thing is clear that cryptocurrencies (and hence Blockchain) will be the matter of discussion for upcoming years and this article may be useful as a reference for further research and studies in the said regard.